\documentclass[proof]{WileyASNA-v1}
\usepackage{supertabular}
\usepackage{lscape}
\articletype{Article Type}%

\received{29 January 2021}
\revised{19 April 2021}
\accepted{03 May 2021}

\begin{document}

\title{Follow-up spectroscopy of comet C/2020\,F3~(NEOWISE)}

\author[]{Richard Bischoff* and Markus Mugrauer}

\authormark{Bischoff \textsc{\& Mugrauer}}

\address[]{\orgdiv{Astrophysikalisches Institut und Universit\"{a}ts-Sternwarte Jena},  \orgaddress{\state{Schillerg\"{a}{\ss}chen 2, 07745 Jena}, \country{Germany}}}

\corres{*R. Bischoff, Astrophysikalisches Institut und Universit\"{a}ts-Sternwarte Jena, Schillerg\"{a}{\ss}chen 2, 07745 Jena, Germany \email{richard.bischoff@uni-jena.de}}

\abstract{We present spectroscopy of the coma center of comet C/2020\,F3~(NEOWISE), carried out at the end of July 2020 with the \'Echelle spectrograph FLECHAS at the University Observatory Jena. The comet was observed in 5 nights and many prominent emission features were detected between 4685\,\AA~ and 7376\,\AA. Beside the $C_2$ Swan emission bands also several emission features of the amidogen radical, as well as two forbidden lines of oxygen were identified in the FLECHAS spectra of the comet in all observing epochs. In contrast, strong sodium emission was detected only in the spectra of the comet, taken on 21 and 23 July 2020, which significantly faded between these two nights, and was no longer present in the spectra as of 29 July 2020. In this paper we present and characterize the most prominent emission features, detected in the FLECHAS spectra of the comet, discuss their variability throughout our spectroscopic monitoring campaign, and use them to derive the radial velocity of the comet in all observing nights.}


\maketitle

\section{Introduction}

Comet C/2020\,F3~(NEOWISE) was discovered on 27 March 2020 by the Wide-Field Infrared Survey Explorer\linebreak (\href{https://minorplanetcenter.net/mpec/K20/K20G05.html}{MPEC\,2020-G05}). This long-period comet revolves around the Sun on a retrograde ($i$$\sim$128.9$^\circ$) highly eccentric ($e$$\sim$0.999) orbit, passed through its perihelion on 3 July 2020 ($q$$\sim$0.295\,au) and reached its closest encounter with the Earth ($\Delta$$\sim$0.692\,au) on 23 July 2020 (orbit reference: JPL{\#}23, as listed in the JPL Small-Body Database\footnote{Online reachable at: \url{https://ssd.jpl.nasa.gov/sbdb.cgi}}). According to International Comet Quarterly\footnote{Online available at: \url{http://www.icq.eps.harvard.edu/index.html}} comet C/2020\,F3~(NEOWISE) was the brightest comet, observable on the sky, in more than a decade, especially since the appearance of comet C/2006 P1 (McNaught) that reached its perihelion in January 2007 or the strong magnitude outburst of comet 17P/Holmes \citep[see e.g.][]{mugrauer2009} end of October of the same year.

In this article we present spectroscopic observations of comet C/2020\,F3~(NEWOWISE), carried out at the University Observatory Jena. In the following section we describe our spectroscopic observations, as well as the data reduction. In section 3 we show the most prominent features detected in the FLECHAS spectra of the comet and analyze their variability throughout our spectroscopic monitoring campaign. Furthermore, we present in this section the measurement of the comet's radial velocity for all observing epochs. Finally, the results of our follow-up spectroscopy of comet C/2020\,F3~(NEWOWISE) are discussed in last section of this paper.

\section{Spectroscopic Observations and Data Reduction}

\begin{figure}[h!]
\includegraphics[width=\linewidth]{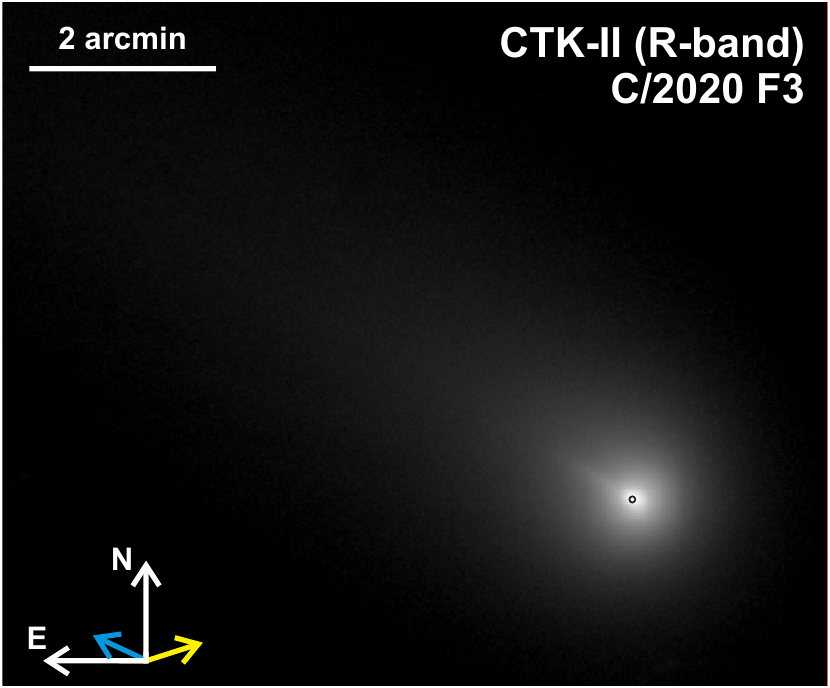}
\caption{The image is the average of ten CTK-II frames, each with an integration time of 10\,s. The size and location of the FLECHAS fiber is indicated as black circle. The blue arrow indicates the direction of the anti-solar vector and the yellow one is the negative heliocentric velocity vector of the comet.}
\label{ctkii}
\end{figure}

Comet  C/2020\,F3 (NEWOWISE) was observed in five nights at the end of July 2020 with the \'Echelle spectrograph FLECHAS \citep{mugrauer2014}, which is operated at the Nasmyth-focus of the 90\,cm-telescope of the University Observatory Jena \citep{pfau}. During the observations the fiber of the spectrograph, which exhibits a diameter of 3.9\,arcsec on the sky, was always guided on the center of the comet's coma, as it is illustrated in Figure\,\ref{ctkii}\hspace{-2mm}. This figure shows a detailed R-band image of the coma and the beginning of the tail of comet C/2020\,F3~(NEWOWISE), taken on 23 July 2020 at 20:30\,UT (mid-time of observation) with the Cassegrain-Teleskop-Kamera \citep[CTK-II, ][]{mugrauer2014}, which is operated at the 25\,cm-Cassegrain telescope of the University Observatory Jena. The shown field of view measures 8.8\,$'\times$7.3\,$'$.

\begin{figure}[h!]
\includegraphics[width=\linewidth]{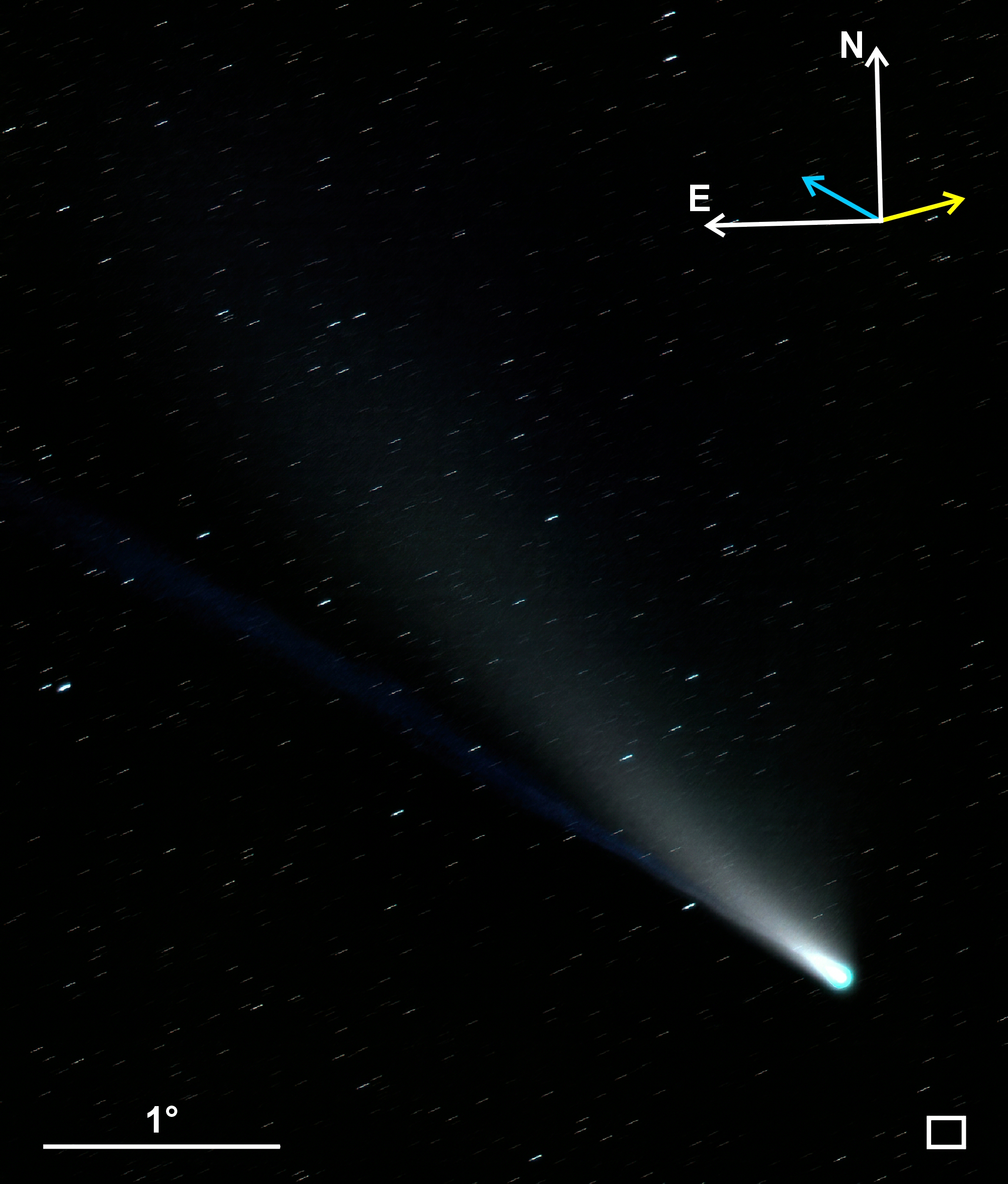}
	\caption{This image is the average of 30 frames, each taken with a detector integration time of 30\,s. The white box in the lower right corner indicates the field of view of the CTK-II image, shown in Figure\,\ref{ctkii}\hspace{-2mm}. The blue arrow indicates the direction of the anti-solar vector and the yellow one is the negative heliocentric velocity vector of the comet.}
	\label{comettail}
\end{figure}

Another image of the comet but with a much wider field of view is shown in Figure\,\ref{comettail}\hspace{-2mm}. This frame shows comet C/2020\,F3~(NEOWISE) with its prominent dust and bluish plasma tail, imaged in the visible spectral range with a Canon\,70\,D digital single-lens reflex camera ~ ($f=200$\,mm, $f/D=2.8$) on 22 July 2020 at 20:34\,UT (mid-time of observation), just 4.4\,hours before the closest encounter of the comet with the Earth. The field of view is $4.25^{\circ}\times5^{\circ}$. The stars appear as trails due to the fast motion of the comet during the integration.


Three spectra of the comet were always taken in each observing night. Between the individual observing epochs the detector integration time and binning had to be increased, due the continuous dimming of the comet by about 2\,mag throughout our spectroscopic monitoring campaign. A detailed overview of the used instrument setup in each night is summarized in the observation log, shown in Table\,\ref{obslog}\hspace{-2mm}. There in we also list the heliocentric distance \textbf{($r$)} of the comet, and its distance to the Earth \textbf{($\Delta$)} for all observing epochs, as calculated with the \texttt{JPL HORIZONS system}\footnote{Online reachable at: \url{https://ssd.jpl.nasa.gov/?horizons}}, using the most recent orbital solution of the comet (orbit reference: JPL{\#}23) based on astrometric measurements.

\begin{table*}[h!]
\caption{Observation log. We list for each observing epoch the mid-time of the spectroscopic observations (Obs-Mid-Time), the airmass \textbf{($X$)} of the target and the altitude of the Sun ($Alt_\odot$) during the observations, as well as the used setup of the spectrograph, namely its detector integration time (DIT), binning mode (binning), resolving power ($R$), as well as the root-mean-square of the obtained wavelength calibration (RMS$_{\text{wavecal}}$). Furthermore, we list the calculated heliocentric distance ($r$) of the comet, and its distance to the Earth ($\Delta$) for each observing epoch.}
\centering
\begin{tabular}{ccccccccc}
\hline
Obs-Mid-Time [Date UT]  & $X$ & $Alt_\odot$ [$^\circ$] & DIT [s]    &  binning & $R$        & RMS$_{\text{wavecal}}$ [m\AA] & $r$ [au] & $\Delta$ [au] \\
\hline
21 July 2020 21:12 & 2.8 & $-$13.4     & 150        & 1$\times$1  & $\sim9300$ & $\sim20$	                    & 0.6040   & 0.6938	\\
23 July 2020 20:30 & 2.2 & $-$9.9\,\,\,& 150        & 1$\times$1  & $\sim9300$ & $\sim20$                      & 0.6481   & 0.6928	\\
29 July 2020 21:34 & 3.1 & $-$16.8     & 300        & 1$\times$1  & $\sim9300$ & $\sim20$	                    & 0.7813   & 0.7537	\\
30 July 2020 21:03 & 2.6 & $-$14.6     & 300        & 2$\times$2  & $\sim7400$ & $\sim40$	                    & 0.8025   & 0.7710	\\
31 July 2020 21:12 & 2.8 & $-$15.6     & 300        & 2$\times$2  & $\sim7400$ & $\sim40$	                    & 0.8241   & 0.7906	\\
\hline
\end{tabular}
\label{obslog}
\end{table*}

FLECHAS is equipped with a back-illuminated CCD-sensor (E2V\,CCD47-10), which consists of 1056\,x\,1027\,pixels, each a square with an edge length of 13\,$\mu$m and covers the spectral range from  3900\,\AA~to 8100\,\AA. The bias level is always corrected by reading out and measuring an overscan region. The typical read-noise of the FLECHAS detector is about 11\,$e^{-}$ and the gain is $1.3\,e^{-}$/ADU. The spectrograph is described in detail in \cite{mugrauer2014}.

For calibration purposes, three frames of a tungsten lamp and three frames of ThAr lamp were always taken for flat-fielding and wavelength calibration, respectively, directly before the observation of the comet. The detector integration time of each calibration image is 5\,s for $1\times 1$-, and 1.5\,s for $2\times 2$-binning, respectively. For dark current- and bias-subtraction, three dark frames were also always recorded for all used integration times, in each observing night.

The data reduction was carried out with the FLECHAS pipeline, which was developed at the Astrophysical Institute Jena. This routine includes dark current subtraction, flat-fielding, the extraction and wavelength calibration of the individual spectral orders, as well as the final averaging and normalization of the reduced spectra \citep{mugrauer2014}.

\section{Variability of Detected Spectral Features and Radial Velocity Measurements}

Many emission features are present in the FLECHAS spectra of comet C/2020\,F3~(NEOWISE), which were taken in the course of our spectroscopic follow-up observing campaign of the comet, which was also announced in \cite{ATel}\footnote{Online available at:\newline \url{https://www.astronomerstelegram.org/?read=13928}}. The individual features were identified by measuring their central wavelengths in the reduced FLECHAS spectra with the IRAF \citep{tody} task \texttt{splot} and comparing them to the reference (laboratory) wavelengths, as listed in the catalogues of cometary emission lines of \cite{brown1996}, and \cite{zhang2001}, respectively. The Swan emission bands of C$_2$ with their band heads at 4685, 4698, 4715, 4737, 5165, 5502, 5541, 5586, 5636, 6005, 6060, 6122, and 6191\,\AA~are well detected, among them the band head at 5165\,\AA, which is the strongest emission feature found in all FLECHAS spectra of the comet. Furthermore, within the wavelength rage between 4925 and 7376\,\AA, numerous emission features of the amidogen radical NH$_{2}$ could be identified in all spectra of the comet, the most prominent ones at 5703, 5732, 5977, 5995, 6020, 6335, and 6641\,\AA, as well as the forbidden lines of oxygen at 6300 and 6364\,\AA. In contrast to these permanently detected emission features a strong sodium emission (line doublet at 5890 \& 5896\,\AA) was present only in the FLECHAS spectra of the comet, taken on 21 and 23 July 2020. The sodium emission significantly faded between these two observing epochs and was not detectable anymore in the FLECHAS spectra of comet C/2020\,F3~(NEOWISE), taken on or after 29 July 2020.

\begin{table}[h!]
\caption{List of all detected emissions features within the spectra of comet C/2020\,F3~(NEOWISE), sorted by their laboratory wavelength.}
\centering
\begin{tabular}{cc}
\hline
feature & feature\\
\hline
$ \text{C}_{2}~\text{band head~(4685)}$ & $\text{C}_{2}~\text{band head~(6060)}$ \\
$ \text{C}_{2}~\text{band head~(4698)}$ & $\text{NH}_{2}~(6097)$                 \\
$ \text{C}_{2}~\text{band head~(4715)}$ & $\text{NH}_{2}~(6098)$                 \\
$ \text{C}_{2}~\text{band head~(4737)}$ & $\text{NH}_{2}~(6109)$                 \\
$ \text{C}_{2}~\text{band head~(5129)}$ & $\text{NH}_{2}~(6110)$                 \\
$ \text{C}_{2}~\text{band head~(5165)}$ & $\text{NH}_{2}~(6121)$                 \\
$\text{NH}_{2}~(5186)$                  & $\text{C}_{2}~\text{band head~(6122)}$ \\
$\text{NH}_{2}~(5194)$                  & $\text{C}_{2}~\text{band head~(6191)}$ \\
$\text{NH}_{2}~(5384)$                  & $\text{NH}_{2}~(6274)$                 \\
$\text{NH}_{2}~(5398)$                  & $\text{NH}_{2}~(6286)$                 \\
$\text{NH}_{2}~(5419)$                  & $\text{NH}_{2}~(6288)$                 \\
$\text{NH}_{2}~(5428)$                  & $\text{NH}_{2}~(6297)$                 \\
$\text{NH}_{2}~(5444)$                  & $[\text{O\,I}]~(6300)$                 \\
$ \text{C}_{2}~\text{band head~(5502)}$ & $\text{NH}_{2}~(6321)$                 \\
$ \text{C}_{2}~\text{band head~(5541)}$ & $\text{NH}_{2}~(6327)$                 \\
$ \text{C}_{2}~\text{band head~(5586)}$ & $\text{NH}_{2}~(6333)$                 \\
$ \text{C}_{2}~\text{band head~(5636)}$ & $\text{NH}_{2}~(6335)$                 \\
$\text{NH}_{2}~(5682)$                  & $\text{NH}_{2}~(6344)$                 \\
$\text{NH}_{2}~(5688)$                  & $\text{NH}_{2}~(6345)$                 \\
$\text{NH}_{2}~(5698)$                  & $\text{NH}_{2}~(6346)$                 \\
$\text{NH}_{2}~(5701)$                  & $\text{NH}_{2}~(6360)$                 \\
$\text{NH}_{2}~(5703)$                  & $\text{NH}_{2}~(6362)$                 \\
$\text{NH}_{2}~(5705)$                  & $ [\text{O\,I}]~(6364)$                \\
$\text{NH}_{2}~(5707)$                  & $\text{NH}_{2}~(6470)$                 \\
$\text{NH}_{2}~(5714)$                  & $\text{NH}_{2}~(6533)$                 \\
$\text{NH}_{2}~(5718)$                  & $\text{NH}_{2}~(6537)$                 \\
$\text{NH}_{2}~(5720)$                  & unidentified line (6578)               \\
$\text{NH}_{2}~(5732)$                  & $\text{NH}_{2}~(6600)$                 \\
$\text{NH}_{2}~(5741)$                  & $\text{NH}_{2}~(6601)$                 \\
$\text{NH}_{2}~(5753)$                  & $\text{NH}_{2}~(6618)$                 \\
$\text{Na}~(5890)$                      & $\text{NH}_{2}~(6619)$                 \\
$\text{Na}~(5896)$                      & $\text{NH}_{2}~(6628)$                 \\
$\text{NH}_{2}~(5963)$                  & $\text{NH}_{2}~(6641)$                 \\
$\text{NH}_{2}~(5977)$                  & $\text{NH}_{2}~(6656)$                 \\
$\text{NH}_{2}~(5985)$                  & $\text{NH}_{2}~(6659)$                 \\
$\text{NH}_{2}~(5995)$                  & $\text{NH}_{2}~(6672)$                 \\
$ \text{C}_{2}~\text{band head~(6005)}$ & $\text{NH}_{2}~(6750)$                 \\
$\text{NH}_{2}~(6007)$                  & $\text{NH}_{2}~(6755)$                 \\
$\text{NH}_{2}~(6019)$                  & $\text{NH}_{2}~(6971)$                 \\
$\text{NH}_{2}~(6020)$                  & $\text{NH}_{2}~(7348)$                 \\
$\text{C}_{2}~\text{band head~(6034)}$  & $\text{NH}_{2}~(7376)$                 \\
\hline
\end{tabular} 		
\label{featurelist}
\end{table}

\begin{table}[h!]
\caption{The average values and standard deviations of the line flux ratios of the selected prominent emission features and the C$_{2}$ emission band head at 5165\,\AA, as measured in all FLECHAS spectra of comet C/2020\,F3~(NEOWISE).}
\centering
\begin{tabular}{cc}
\hline
feature & $\mathcal{F}[\text{feature}]/\mathcal{F}[\text{C}_{2}~(5165)]$ \\
\hline
$ \,\,\,\,\,\text{C}_{2}~\text{band head~(5635)}$ &  $0.37\pm0.07$ \\
\hline
$\text{NH}_{2}~(5703)$  & $0.20\pm0.04$ \\
$\text{NH}_{2}~(5732)$  & $0.24\pm0.08$ \\
$\text{NH}_{2}~(5741)$  & $0.16\pm0.05$ \\
$\text{NH}_{2}~(5753)$  & $0.10\pm0.02$ \\
$\text{NH}_{2}~(5977)$  & $0.53\pm0.06$ \\
$\text{NH}_{2}~(5985)$  & $0.10\pm0.02$ \\
$\text{NH}_{2}~(5995)$  & $0.37\pm0.06$ \\
$\text{NH}_{2}~(6020)$  & $0.19\pm0.02$ \\
$\text{NH}_{2}~(6098)$  & $0.17\pm0.06$ \\
$\text{NH}_{2}~(6335)$  & $0.23\pm0.06$ \\
$\text{NH}_{2}~(6641)$  & $0.26\pm0.06$ \\
\hline
$ [\text{O\,I}]~(6300)$ & $0.32\pm0.09$ \\
$ [\text{O\,I}]~(6364)$ & $0.11\pm0.03$ \\
\hline
\end{tabular} 		
\label{ratios}
\end{table}
\begin{table}[h!]
\caption{The line flux ratios of the detected sodium emission lines and the strong C$_{2}$ band head at 5165\,\AA, as measured in the FLECHAS spectra of comet C/2020\,F3~(NEOWISE).}
\centering
\begin{tabular}{ccc}
\hline
feature & JD-2450000 & $\mathcal{F}[\text{feature}]/\mathcal{F}[\text{C}_{2}~(5165)]$  \\
\hline
Na$_{\text{D2}}$\,(5890)   & 9052.383449 & $0.44\pm0.01$ \\
			& 9054.354444 & $0.15\pm0.01$ \\
\hline
Na$_{\text{D1}}$\,(5896)   & 9052.383449 & $0.29\pm0.01$ \\
			& 9054.354444 & $0.10\pm0.01$ \\
\hline
\end{tabular}  		
\label{ratioNa}
\end{table}
\begin{table}[h!]
\caption{The measured radial velocities \textbf{($RV$)} of comet C/2020\,F3~(NEOWISE) for all observing epochs together with the radial velocities \textbf{($RV_{\text{calc}}$)}, calculated with the \texttt{JPL HORIZON system} for the observing site, based on the most recent astrometric orbital solution of the comet.}
\centering
\begin{tabular}{ccc}
\hline
 JD-2450000 & $RV$ [km/s] & $RV_{\text{calc}}$ [km/s]\\
\hline
9052.383449 & \,\,\,$-6.8\pm1.3$ & $-5.8$\,\,\,\\
9054.354444 & \,\,\,$+2.5\pm1.3$ & $+4.2$\,\,\,\\
9060.398333 & $+27.6\pm1.3$& $+29.4$\\
9061.376829 & $+32.9\pm2.4$& $+32.5$\\
9062.383229 & $+35.2\pm2.3$& $+35.3$\\
\hline
\vspace{2mm}
\end{tabular} 		
\label{RV}
\end{table}

In order to characterize the variability of the detected most prominent emission features within the span of time, covered by our spectroscopic monitoring campaign,  we have measured with \texttt{splot} the line fluxes \textbf{($\mathcal{F}$)} of all these features in the individual FLECHAS spectra. Thereby we have selected the strong C$_{2}$ band head at 5165\,\AA\,\,as reference for the calculation of the line flux ratios $\mathcal{F}[\text{feature}]/\mathcal{F}[\text{C}_{2}~(5165)]$, whose average values and standard deviations are summarized in Table\,\ref{ratios}\hspace{-2mm}.

\begin{figure}[h!]\vspace{4.9mm}
\includegraphics[width=\linewidth]{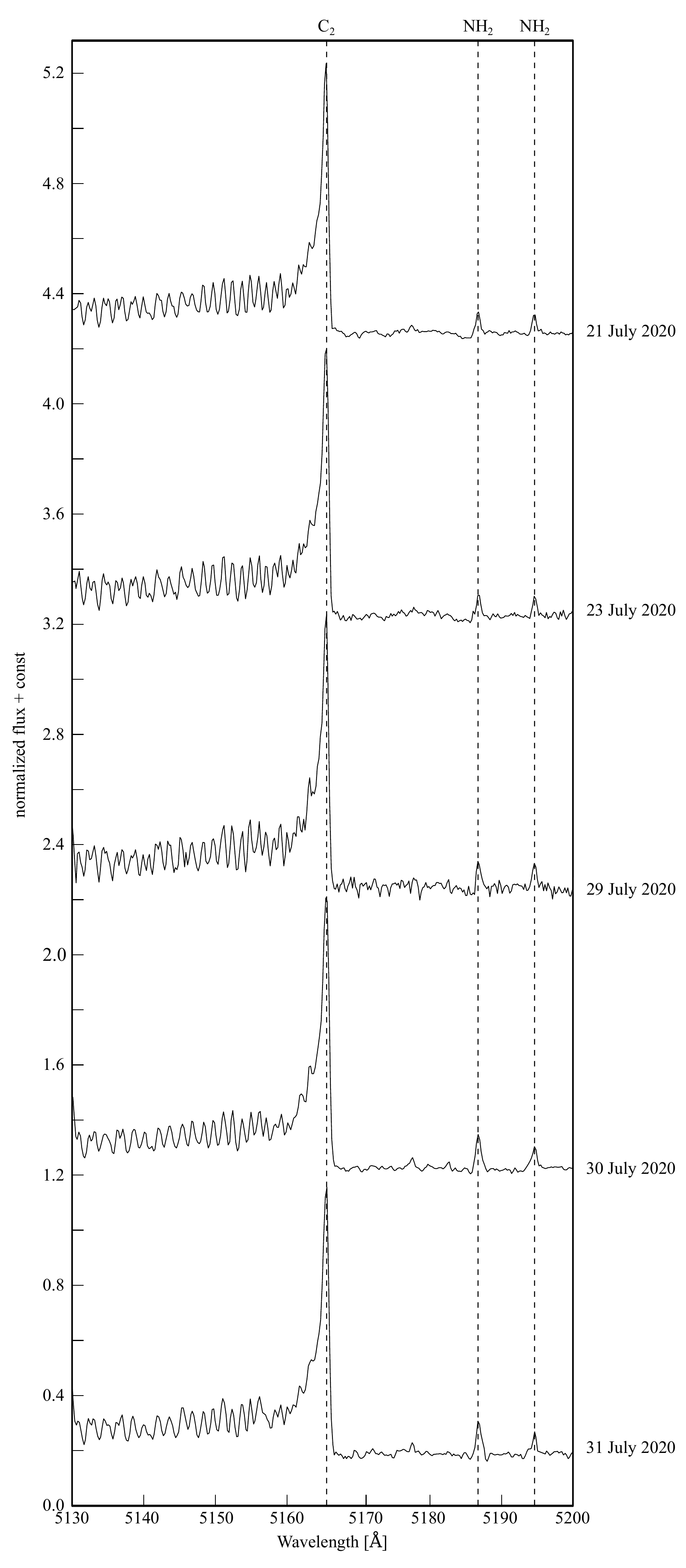}
\caption{Spectra of comet C/2020\,F3~(NEOWISE), normalized to the line flux density of the C$_{2}$ band head at 5165\,\AA.\vspace{10mm}}
\label{ap14}
\end{figure}

\begin{figure}[h!]
\includegraphics[width=\linewidth]{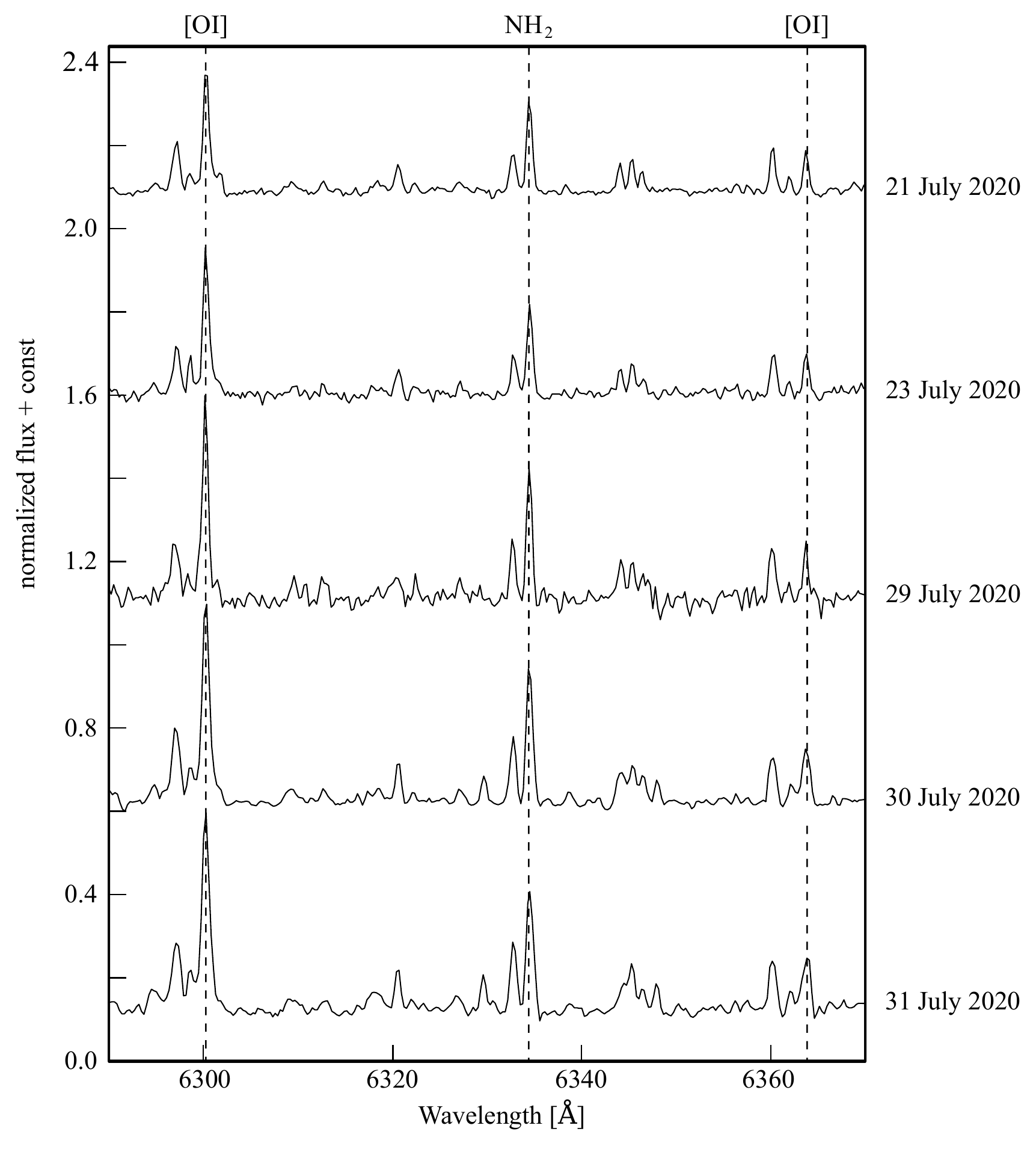}
	\caption{Spectra of the comet C/2020\,F3~(NEOWISE), normalized to the line flux density of the C$_{2}$ band head at 5165\,\AA.}
	\label{ap22}
\end{figure}

The individual spectral ranges, which contain the selected emission features, are illustrated for all observing epochs in the Figures\,\ref{ap14}\hspace{-2mm} to \ref{ap22}\hspace{-2mm}, which are normalized to the line flux density of the most prominent C$_2$ band head at 5165\,\AA.\,\,A list of all detected emission features, detected in the FLECHAS spectra of the comet, taken on the individual observing dates, is given in Table\,\ref{featurelist}\hspace{-2mm}. All listed emission lines were found in every observing night, except the sodium doublet.

\begin{figure}[h!]
\includegraphics[width=\linewidth]{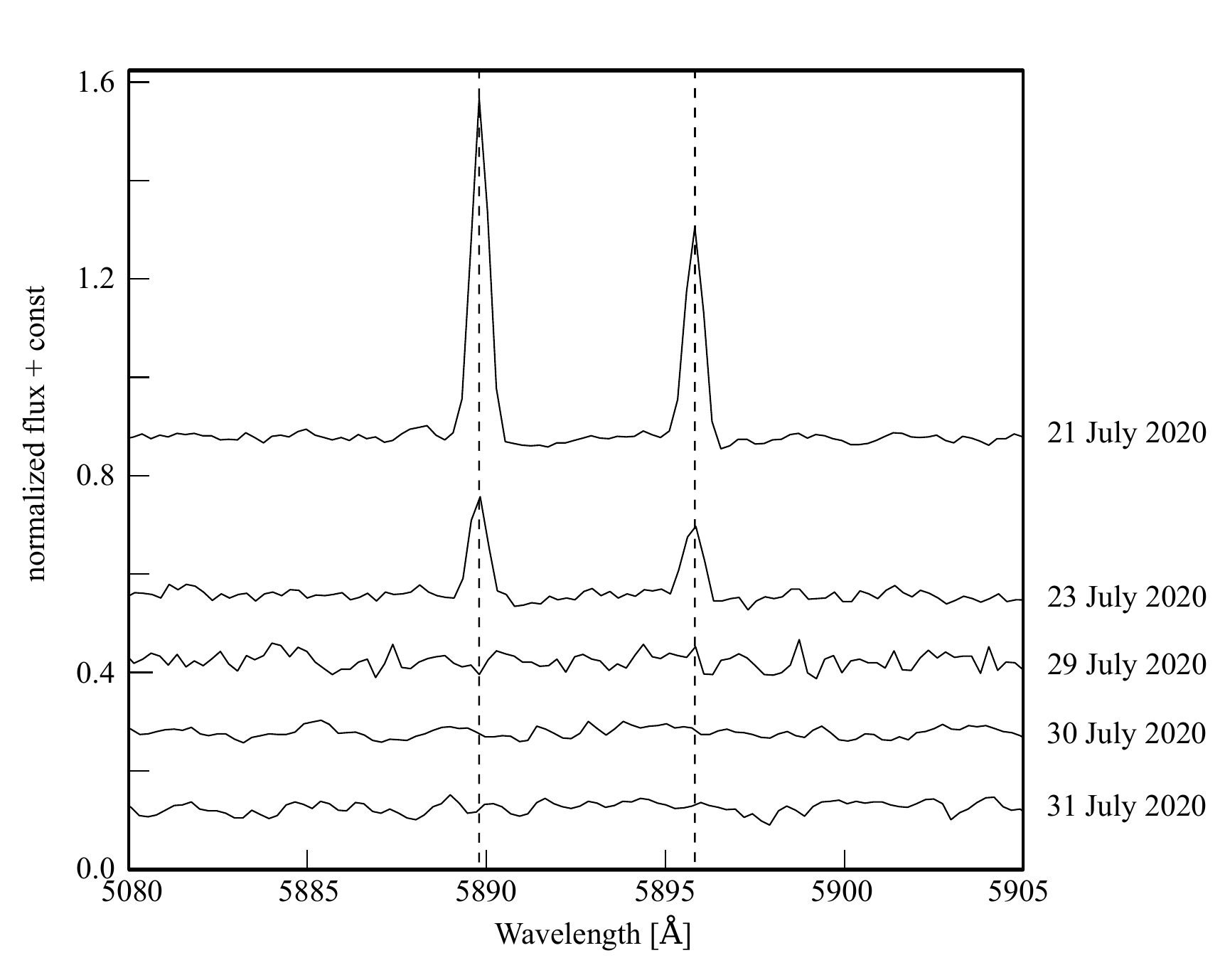}
\caption{The sodium doublet, detected in the FLECHAS spectra of comet C/2020\,F3~(NEOWISE), normalized to the line flux density of the C$_{2}$ band head at 5165\,\AA.}
\label{Na}
\end{figure}

The evolution of the sodium emission lines over all observing epochs is shown in Figure\,\ref{Na}\hspace{-2mm} and the corresponding line flux ratios are summarized in Table\ref{ratioNa}\hspace{-2mm}.

Furthermore, we have also determined the radial velocity of the comet in each observing epoch, by measuring the wavelength shift \textbf{($\Delta \lambda$)} of the detected prominent NH$_{2}$ emission features, listed in Table\,\ref{ratios}\hspace{-2mm}, and in the first two observing epochs also of the sodium doublet, to the corresponding laboratory wavelengths \textbf{($\lambda_0$)}:
\begin{align}
RV=\frac{\Delta \lambda}{\lambda_0} \cdot c
\end{align}

Our results are summarized in Table\,\ref{RV}\hspace{-2mm} together with the radial velocities of the comet, as calculated with the \texttt{JPL HORIZON system} for the observing site, using the most recent orbital solution of the comet (orbit reference: JPL{\#}23), based on astrometric measurements.

\section{Discussion}

In our observing campaign of comet C/2020\,F3~(NEOWISE), which was carried out with FLECHAS at the University Observatory Jena at the end of July 2020, we could take spectra of the comet in five observing nights.

We have measured the line fluxes of the detected most prominent emission features in the comet's spectra, namely those of C$_2$, NH$_2$, oxygen, and sodium. In all spectra the line fluxes were compared to that of the strongest detected emission feature, the {C}$_{2}$ band head at 5165\,\AA, and the corresponding line flux ratios were determined. The obtained ratios, which are summarized in Table\,\ref{ratios}\hspace{-2mm} are stable on the few percent level (5\,\% on average). In contrast, the sodium emission, detected in the first observing epoch on 21 July 2020 significantly faded within about 2 days by a factor of $2.9\pm0.2$ and was not detectable anymore in the spectrum of the comet as on 29 July 2020. Assuming a linear decrease in the sodium emission within the time span, covered by our observations, we find that the sodium emission in the spectrum of comet C/2020\,F3~(NEOWISE) should have disappeared at $JD=2459055.41\pm0.21$, i.e. when the comet was at a heliocentric distance of about 0.67\,au on its trajectory away from the Sun.

Sodium emission was found so far in the spectra of many in particular bright comets, most recently in those of comet C/2006\,P1~(McNaught), comet C/2011\,L4~(PANSTARRS), and the disintegrated comet C/2012\,S1~(ISON) \citep{snodgrass2008, fulle2013, schmidt2015}. Dust particles have been suggested as the main source of sodium in the comae and dust tails of comets. But even after more than one century after the first detection of sodium emission in cometary spectra, the exact mechanisms by which this alkali metal is released from the nucleus and from the dust particles is still under debate \citep{cremonese2002, leblanc2008}. Sodium emission is typically observed in the spectra of bright comets when they are within a heliocentric distance of about 0.8\,au \citep{swamy2010}. According to the results of our FLECHAS spectroscopy, presented here, this is also the case for comet C/2020\,F3~(NEOWISE).

A detailed description of the composition of cometary volatiles is given in \cite{book} and \cite{book2}.

Beside the analysis of the variability of prominent emission features, detected in the spectrum of comet C/2020\,F3~(NEOWISE), we also have determined the radial velocity of the comet in each observing epoch by measuring the wavelengths of the strong emission features of NH$_2$ and in the first two observing nights also of sodium. The obtained radial velocities are summarized in Table\,\ref{RV}\hspace{-2mm} together with the radial velocities, calculated with the \texttt{JPL HORIZON system}, using the most recent orbital solution of the comet (orbit reference: JPL{\#}23), based on astrometric measurements. The radial velocity of the comet is negative only in the first observing night, because the comet was observed then just about 27.8 hours before its closest encounter with the Earth. In the following observing epochs the radial velocity of comet C/2020\,F3~(NEOWISE) significantly increased as it moved away from the Earth. The obtained radial velocity measurements, derived from our FLECHAS spectroscopy, are well consistent with the calculated radial velocities of the comet (root-mean-square of 1.2\,km/s), based on its astrometric orbit.

\bibliography{bischoff}\vspace{5mm}

\begin{figure}[h!]
\includegraphics[width=\linewidth]{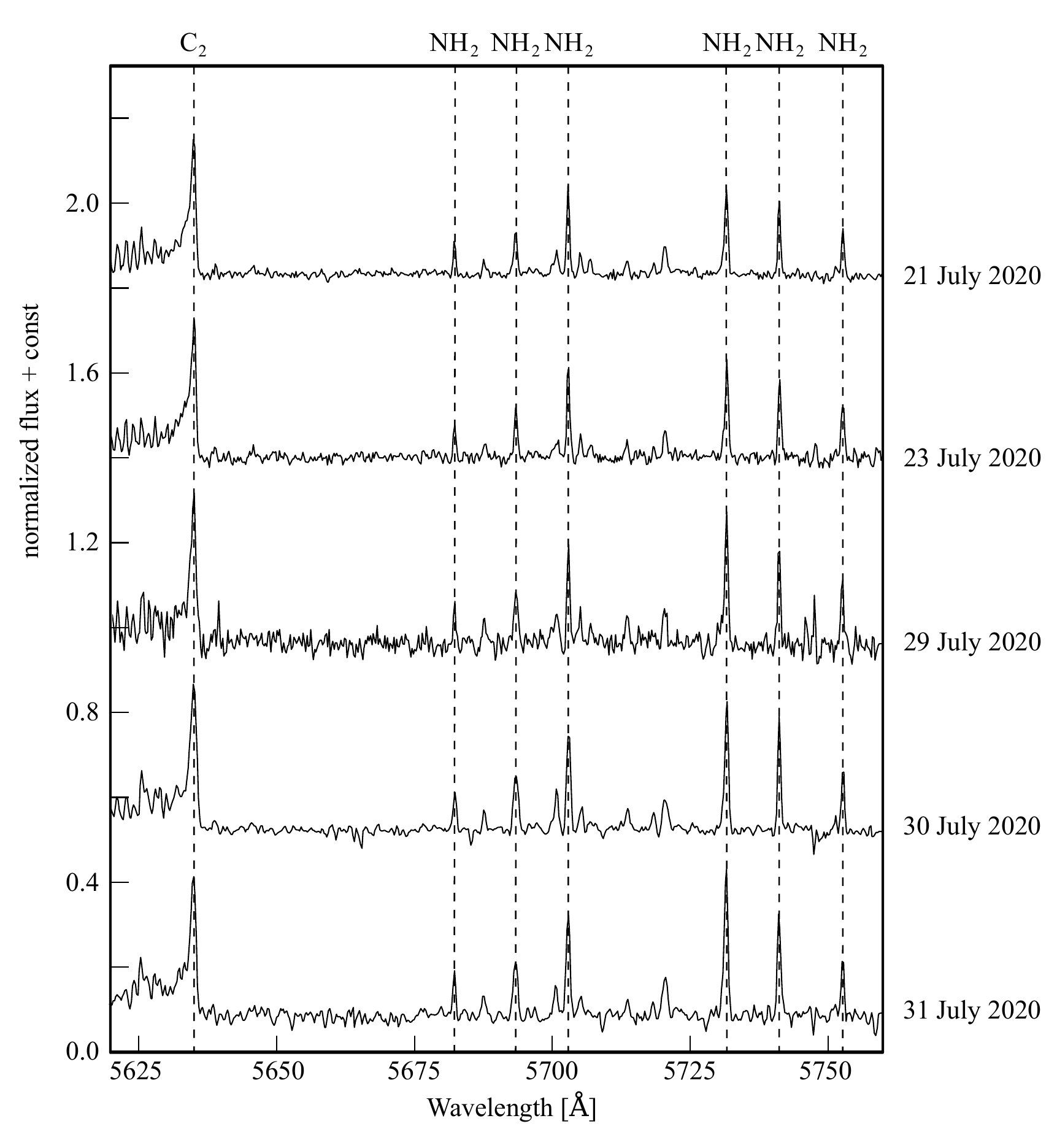}
	\caption{Spectra of comet C/2020\,F3~(NEOWISE), normalized to the line flux density of the C$_{2}$ band head at 5165\,\AA.}
	\label{ap18}
\end{figure}
\begin{figure}[h!]
\includegraphics[width=\linewidth, height=9.3cm]{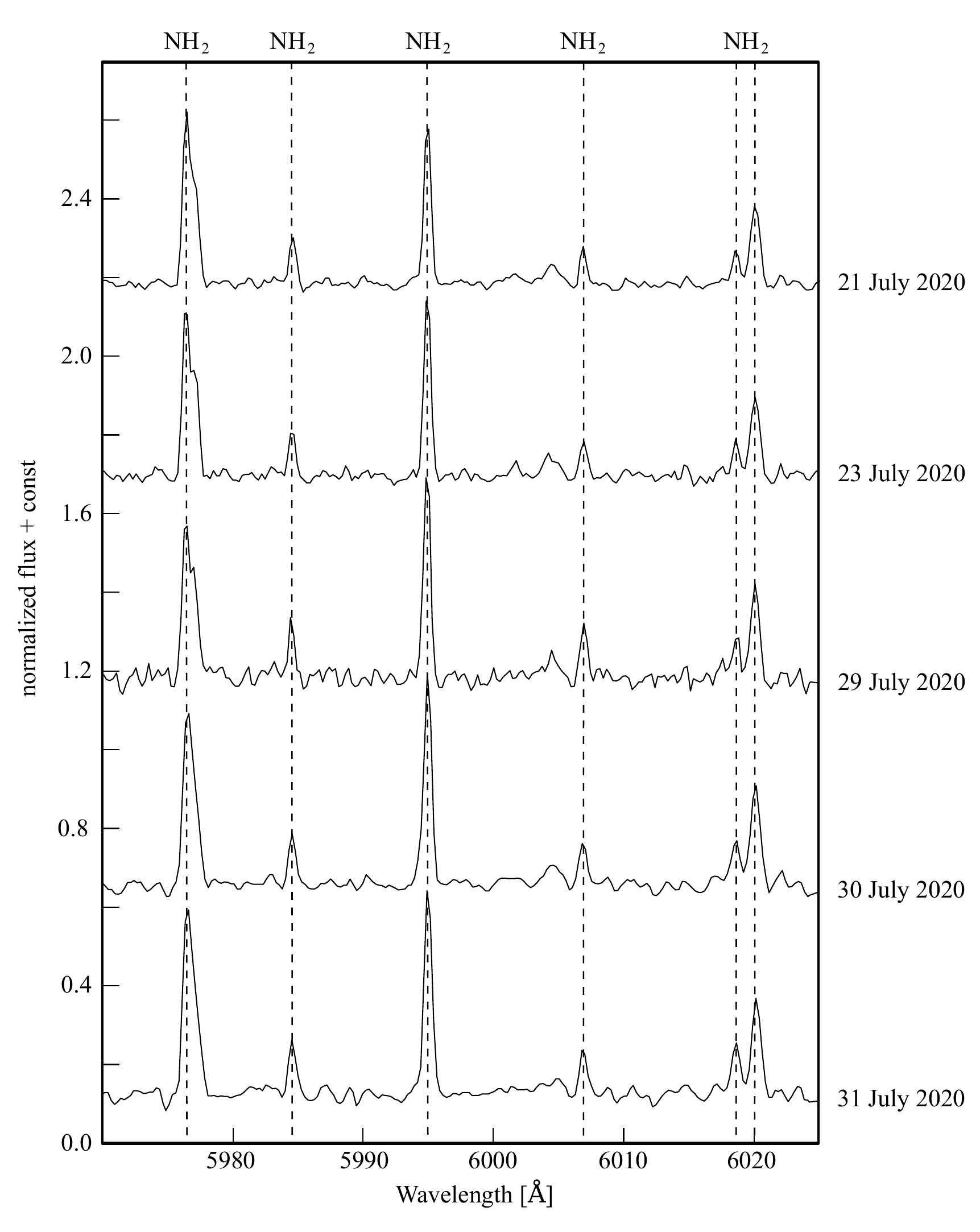}
	\caption{Spectra of comet C/2020\,F3~(NEOWISE), normalized to the line flux density of the C$_{2}$ band head at 5165\,\AA.}
	\label{ap20}
\end{figure}


\section*{Acknowledgments}

This work was supported by the Deutsche Forschungsgemeinschaft with the projects \fundingNumber{NE 515/58-1} and \fundingNumber{MU 2695/27-1}.

\section*{Author Biography}

Richard Bischoff is a PhD student at the Astrophysical Institute and University Observatory Jena. His main fields of research are photometry and spectroscopy of exoplanet candidate host stars.

\end{document}